\documentclass[mlabstract]{jmlr}

\usepackage{longtable}

\usepackage{booktabs}
\usepackage[load-configurations=version-1]{siunitx} 

\theorembodyfont{\upshape}
\theoremheaderfont{\scshape}
\theorempostheader{:}
\theoremsep{\newline}

\jmlrvolume{}
\firstpageno{1}
\editors{List of editors' names}

\jmlryear{2026}
\jmlrworkshop{Symmetry and Geometry in Neural Representations}

\usepackage{multirow}

\usepackage{xcolor}
\usepackage[dvipsnames]{xcolor}

\definecolor{speedupgreen}{RGB}{0,135,70}
\definecolor{slowdownred}{RGB}{155,55,50}

\usepackage{tikz}
\usepackage{cleveref}
\usetikzlibrary{arrows.meta,calc,positioning}

\definecolor{scoregood}{HTML}{4C956C}
\definecolor{scorelow}{HTML}{BC4749}
\definecolor{plotink}{HTML}{202A34}
\definecolor{plotpanel}{HTML}{F1F3F5}

\crefname{figure}{Fig.}{Figs.}
\Crefname{figure}{Fig.}{Figs.}

\crefname{table}{Tab.}{Tabs.}
\Crefname{table}{Tab.}{Tabs.}

\crefname{section}{Sec.}{Secs.}
\Crefname{section}{Sec.}{Secs.}
\newcommand{\lmax}
{\ensuremath{\ell_{\max}}}

\newcommand{\lthree}{\ensuremath{\ell_{\max}\mathord{=}3}}
\newcommand{\lfour}{\ensuremath{\ell_{\max}\mathord{=}4}}

\newcommand{\nthree}{\ensuremath{n_{\max}\mathord{=}3}}
\newcommand{\nfive}{\ensuremath{n_{\max}\mathord{=}5}}
\newcommand{\nfour}{\ensuremath{n_{\max}\mathord{=}4}}
\newcommand{\nseven}{\ensuremath{n_{\max}\mathord{=}7}}

\title[Short Title]{Rotation-Invariant Machine Learning for \\Highly Symmetric Molecular Configurations}

\author{%
  \Name{Logan Bolton} \\ 
    \addr CAI-3: Information Sciences, Los Alamos National Laboratory
     \AND
  \Name{Tomas Karella\nametag{\thanks{Corresponding author}}} \Email{karella@lanl.gov}\\
    \addr CAI-3: Information Sciences, Los Alamos National Laboratory
     \AND
  \Name{Nicholas Lubbers} \\
    \addr CAI-3: Information Sciences, Los Alamos National Laboratory
     \AND
  \Name{Nathaniel Philip  Gorski} \\
    \addr CAI-3: Information Sciences, Los Alamos National Laboratory
     \AND
  \Name{Emily Shinkle} \\
  \addr CAI-3: Information Sciences, Los Alamos National Laboratory
  \AND
  \Name{Alice Elisabeth Anastasia Allen} \\
  \addr Machine Learning in Molecular Dynamics, Max Planck Institute for Polymer Research
  \AND
  \Name{Pieter J Swart} \\
  \addr T-5: Applied Mathematics and Plasma Physics, Los Alamos National Laboratory
  \AND
  \Name{Roxana Bujack} \\
  \addr CAI-3: Information Sciences, Los Alamos National Laboratory
}

\begin{document}

\maketitle


\begin{abstract}

Atomistic machine learning can approximate expensive quantum mechanical calculations at a fraction of their computational cost, but accuracy depends strongly on how the three-dimensional atomic environments are represented. Invariant representations enable fast computation but can fail to distinguish geometrically distinct configurations, especially in highly symmetric environments. We introduce a lightweight benchmark to evaluate expressivity on $n$-fold symmetric configurations, and use it to expose degeneracies in both invariant and equivariant architectures. Guided by these results, we construct a new invariant basis for HIP-HOP which resolves most of the identified failures. 
On a methane simulation dataset, the extended basis achieves the lowest error among the HIP-HOP variants and has similar accuracy to EquiformerV3 while running $3.2\times$ faster at inference.
These results show that carefully constructed invariant representations can mitigate key expressivity limitations while balancing accuracy and computational efficiency.
\end{abstract}


\section{Introduction}
Accurate atomistic simulation is central to understanding material properties, chemical reactions, and biological processes. While first principles quantum mechanical (QM) calculations~\citep{hohenberg1964,levy1979,vignale1987} provide highly accurate results, their computational cost makes routine large-scale and long-timescale simulations challenging. Machine-learned interatomic potentials (MLIPs) have therefore emerged as efficient surrogates, reproducing QM-level accuracy at a fraction of the cost~\citep{behler-parrinello,smith2017}. Recent large-scale models such as UMA demonstrate how far this paradigm has advanced, scaling to hundreds of millions of training structures while enabling fast inference across diverse chemical systems~\citep{wood2026}.

A key factor behind the success of MLIPs is the incorporation of physical symmetries as inductive biases. For example, energy predictions are invariant to translations, rotations, and permutations of identical atoms, while forces transform along with the geometric transformations. Encoding these properties directly into neural architectures has become a cornerstone of geometric deep learning~\cite{bronstein2021, cohen2016, weiler2023}, leading to substantial gains in accuracy and data efficiency.

These approaches can be broadly divided into two strategies for incorporating symmetry, each with different tradeoffs. Invariant models~\citep{schutt2018, gasteiger2020, wang2022, allen2026optimal} use scalar geometric quantities such as distances, angles, and inner products, enabling highly efficient computation in a reduced representation space, but potentially limiting expressivity if the invariants are not sufficiently rich to describe geometry. In contrast, equivariant models~\citep{equiformer_v3, kondor2018, thomas2018, weiler2018, Batatia2022mace, Batatia2022Design} retain vector and tensor-valued features that transform consistently under symmetry operations, preserving directional information and improving expressivity. This comes at increased computational cost due to operations on multi-component tensor representations. 

This raises a central question: is explicit equivariant machinery strictly necessary for all tasks? Invariant representations are not inherently lossy; rather, their adequacy depends on whether they are complete, i.e., whether they distinguish all molecular configurations up to the prescribed symmetries. This suggests an alternative design point, namely constructing complete invariant representations that retain all symmetry-relevant geometric information while enabling fast computation.

In this work, we introduce a new basis for HIP-HOP, an invariant architecture combining completeness with computational efficiency~\citep{allen2026optimal}. Using an angular expansion up to \lfour, the representation captures finer angular structure while retaining invariance. The extended basis resolves the identified degeneracies and improves predictive accuracy. 
This new model has similar accuracy to EquiformerV3 while offering higher inference throughput, though EquiformerV3 remains more accurate at larger model sizes.

\section{Method}
We interpret the local atomistic configuration as a density on a
sphere and describe its shape by projecting it onto a polynomial basis truncated at a maximal angular order $\ell_{\max}$, following the theory
of moment invariants~\citep{1057692} in the flexible formulation of \cite{bujack2025}. Intuitively, $\ell_{\max}$ plays the same role as the truncation order of a Fourier or Taylor series. Low orders capture only coarse angular features of the neighbor distribution, while higher orders resolve increasingly fine angular detail at increased computational cost. This mirrors the angular truncation used in atom-centered descriptors such as  ACE~\citep{PhysRevB.99.014104}, and in equivariant networks built on spherical harmonics~\citep{thomas2018}. The resulting projections, called geometric moments, form Cartesian tensors. Because products and contractions of tensors are again tensors, every rank-zero contraction of a product of moment tensors is a scalar and therefore a rotation invariant. Infinitely many such contractions exist and most are redundant, so we identify an independent subset using Langbein's algorithm~\citep{ langbein2014, bujack2025} to obtain a compact basis.

\paragraph{Benchmarks}
We first evaluate HIP-HOP on the N-body benchmark~\citep{pozdnyakov2020, joshi2023expressive} to compare the ability of invariant and equivariant models to resolve geometric interactions. To complement the N-body benchmark, we introduce a lightweight synthetic dataset designed to evaluate highly symmetric local environments, including configurations with $n$-fold symmetry. This benchmark provides a controlled setting that is less contrived than the N-body benchmark while remaining simple enough to isolate differences in model expressivity, and is motivated by known degeneracies of finite-order atom-centered invariant representations \citep{bujack2025}.
We construct graphs with a central node surrounded by two rings and vary their populations and relative rotation to generate configurations with different $n$-fold symmetries. For each configuration, we consider orientations that minimize or maximize the distance between nodes on different rings, as illustrated in \cref{fig:ring-rotation}. Additional details can be found in Appendix~\ref{apd:ring-details}.

\begin{figure}
    \centering
    \resizebox{0.6\linewidth}{!}{%
        \begin{minipage}{\linewidth}
            \centering

            \begin{minipage}[c]{0.27\linewidth}
                \centering
                \includegraphics[
                    width=0.8\linewidth,
                    trim=0 0 0 0,
                    clip
                ]{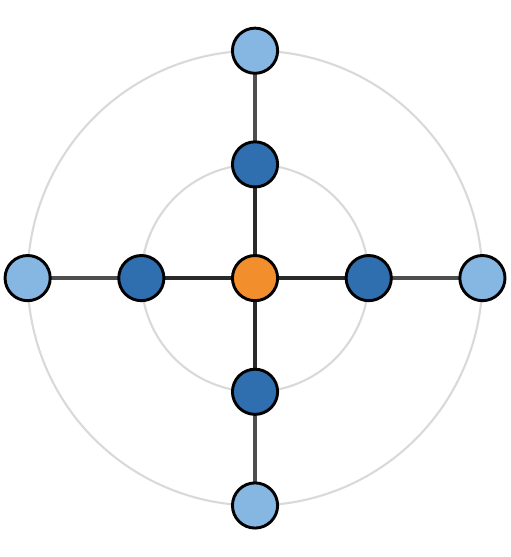}

                \vspace{-0.4em}

                {\large \textit{Class 1: Close}}
            \end{minipage}
            \hspace{0.06\linewidth}
            %
            \raisebox{0.5em}{%
                \begin{minipage}[c]{0.16\linewidth}
                    \centering
                    \includegraphics[
                        width=0.8\linewidth,
                        trim=0 0 0 0,
                        clip
                    ]{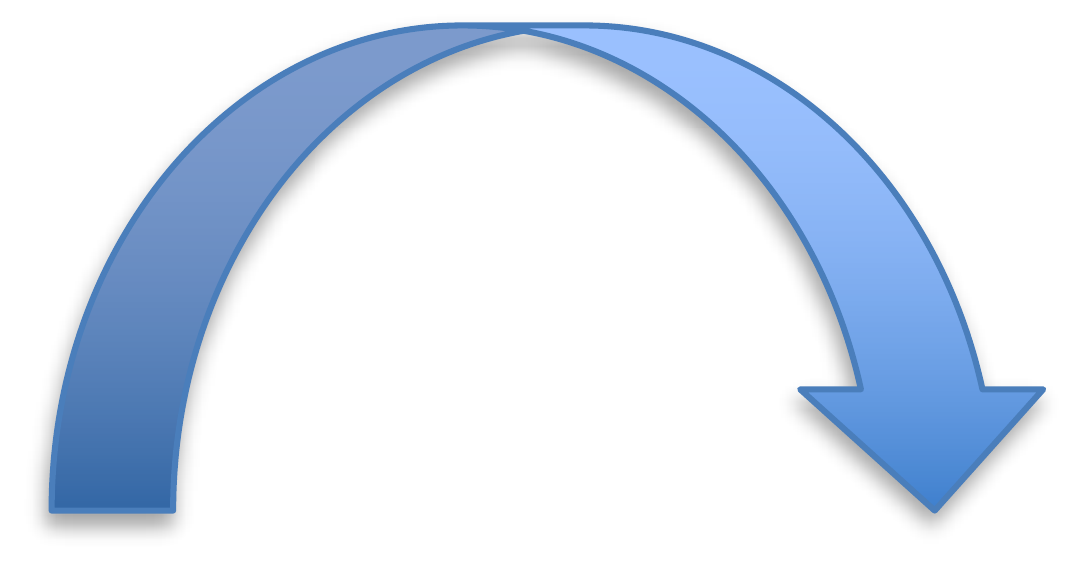}

                    \vspace{0.2em}

                    \makebox[\linewidth][c]{%
                        \large\textbf{Rotated outer ring}%
                    }
                \end{minipage}
            }
            \hspace{0.06\linewidth}
            %
            \begin{minipage}[c]{0.27\linewidth}
                \centering
                \includegraphics[
                    width=0.8\linewidth,
                    trim=0 0 0 0,
                    clip
                ]{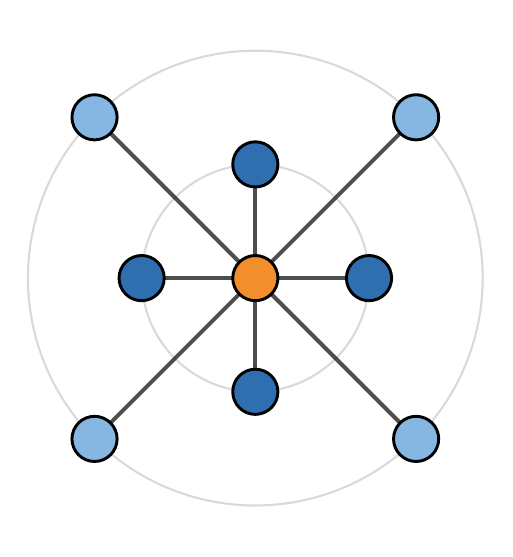}

                \vspace{-0.4em}

                {\large \textit{Class 2: Far}}
            \end{minipage}

        \end{minipage}%
    }

    \caption{For the rings benchmark, the model is given two rotated versions of the same graph, and must determine which orientation has the nodes closer together.}
    \label{fig:ring-rotation}
    \vspace{-0.5em}
\end{figure}

\paragraph{Methane Dataset}
To test how basis expressivity affects real-world performance, we evaluate our model on the randomly displaced methane dataset of \cite{pozdnyakov2020Methan}, containing over seven million configurations with DFT energies and forces. Methane provides a challenging four-neighbor environment, making it a natural test of the additional geometric resolution introduced by extending the basis to \lfour.


\paragraph{Model}
We extend the original construction of HIP-HOP using the flexible moment-invariant framework of \cite{bujack2025} to obtain a new basis through \lfour. The resulting basis and its tensor-contraction structure are shown in Appendix~\ref{app:basis}. Increasing the basis size improves model expressivity but also introduces higher-order polynomial terms and increased computational cost. To mitigate this overhead, we parallelize both the construction and evaluation of the invariant polynomials.
\vspace{-0.5em}
\section{Results}


\definecolor{scoregood}{RGB}{47,128,55}
\definecolor{scorelow}{RGB}{207,45,45}
\definecolor{plotink}{RGB}{35,35,35}

\begin{figure}
\centering

\resizebox{0.7\linewidth}{!}{%
\begin{tikzpicture}[
    x=1cm,
    y=1cm,
    font=\sffamily
]

\def\cell{0.81}
\def\gridline{0.55pt}
\def\leftlabelx{-0.15}
\def\panelgap{0.1}
\def\rightshift{4*\cell+\panelgap}
\def\groupgap{0.14} 

\newcommand{\drawleftrow}[3]{%
    \foreach \score [count=\column from 0] in {#3}{%
        \ifdim\score pt>0.75pt
            \def\cellcolor{scoregood}%
        \else
            \def\cellcolor{scorelow}%
        \fi

        \fill[\cellcolor]
            ({\column*\cell},{#1-\cell/2})
            rectangle
            ({(\column+1)*\cell},{#1+\cell/2});

        \draw[white, line width=\gridline]
            ({\column*\cell},{#1-\cell/2})
            rectangle
            ({(\column+1)*\cell},{#1+\cell/2});

        \node[
            text=white,
            font=\scriptsize\bfseries
        ] at ({(\column+0.5)*\cell},#1)
            {\score};
    }

    \node[
        anchor=east,
        text=plotink,
        font=\scriptsize
    ] at (\leftlabelx,#1)
        {#2};
}

\newcommand{\drawrightrow}[2]{%
    \foreach \score [count=\column from 0] in {#2}{%
        \ifdim\score pt>0.75pt
            \def\cellcolor{scoregood}%
        \else
            \def\cellcolor{scorelow}%
        \fi

        \fill[\cellcolor]
            ({\rightshift+\column*\cell},{#1-\cell/2})
            rectangle
            ({\rightshift+(\column+1)*\cell},{#1+\cell/2});

        \draw[white, line width=\gridline]
            ({\rightshift+\column*\cell},{#1-\cell/2})
            rectangle
            ({\rightshift+(\column+1)*\cell},{#1+\cell/2});

        \node[
            text=white,
            font=\scriptsize\bfseries
        ] at ({\rightshift+(\column+0.5)*\cell},#1)
            {\score};
    }
}

\def\rowA{0}
\def\rowB{-1*\cell}
\def\rowC{-2*\cell-\groupgap}
\def\rowD{-3*\cell-\groupgap}
\def\sepy{-1.5*\cell-0.5*\groupgap}

\drawleftrow{\rowA}
{EquiformerV3 (\lthree)}
{1.0,1.0,1.0}

\drawleftrow{\rowB}
{HIP-HOP$^*$ (\lthree)}
{1.0,1.0,1.0}

\drawleftrow{\rowC}
{EquiformerV3 (\lfour)}
{1.0,1.0,1.0}

\drawleftrow{\rowD}
{HIP-HOP (\lfour)}
{1.0,1.0,1.0}

\drawrightrow{\rowA}
{1.0,0.5,1.0,0.5,0.5,1.0,0.5,0.5}

\drawrightrow{\rowB}
{1.0,0.5,1.0,0.5,0.5,1.0,0.5,0.5}

\drawrightrow{\rowC}
{1.0,1.0,1.0,1.0,1.0,1.0,0.5,1.0}

\drawrightrow{\rowD}
{1.0,1.0,1.0,1.0,1.0,1.0,0.5,1.0}

\draw[
    draw=plotink,
    line width=0.45pt,
    opacity=0.0,
    dash pattern=on 3.5pt off 2.5pt
] ({\leftlabelx-1.45}, {\sepy}) -- ({\rightshift+8*\cell}, {\sepy});

\node[
    text=plotink,
    font=\scriptsize,
    rotate=40,
    anchor=north east
] at (0.5*\cell,{-3.45*\cell-\groupgap})
{2 body};

\node[
    text=plotink,
    font=\scriptsize,
    rotate=40,
    anchor=north east
] at (1.5*\cell,{-3.45*\cell-\groupgap})
{3 body};

\node[
    text=plotink,
    font=\scriptsize,
    rotate=40,
    anchor=north east
] at (2.5*\cell,{-3.45*\cell-\groupgap})
{\shortstack[r]{4 body\\[-3.05pt]}};


\foreach \inner/\outer [count=\column from 0] in {
    1/3,
    1/4,
    2/2,
    2/3,
    2/4,
    3/3,
    3/4,
    4/4
}{%
    \node[
        text=plotink,
        font=\scriptsize,
        rotate=40,
        anchor=north east
    ] at ({\rightshift+(\column+0.5)*\cell},{-3.45*\cell-\groupgap})
        {$\left(\inner,\outer\right)$};
}

\node[
    text=plotink,
    font=\scriptsize,
    align=center
] at ({\rightshift+4*\cell},{-5.25*\cell-\groupgap})
{\textbf{B)} Rings Benchmark (inner-ring nodes, outer-ring nodes)};

\node[
    text=plotink,
    font=\scriptsize,
    align=center
] at ({1.25*\cell},{-5.25*\cell-\groupgap})
{\textbf{A)} N-body Benchmark};

\end{tikzpicture}
}
\vspace{-0.75em}

\caption{
The $n$-body benchmark (A) accuracy is saturated for all bases, but the rings benchmark (B) shows the difference in expressivity between bases.
\vspace{-1.5em}
}
\label{fig:ring-nbody-results}
\end{figure}
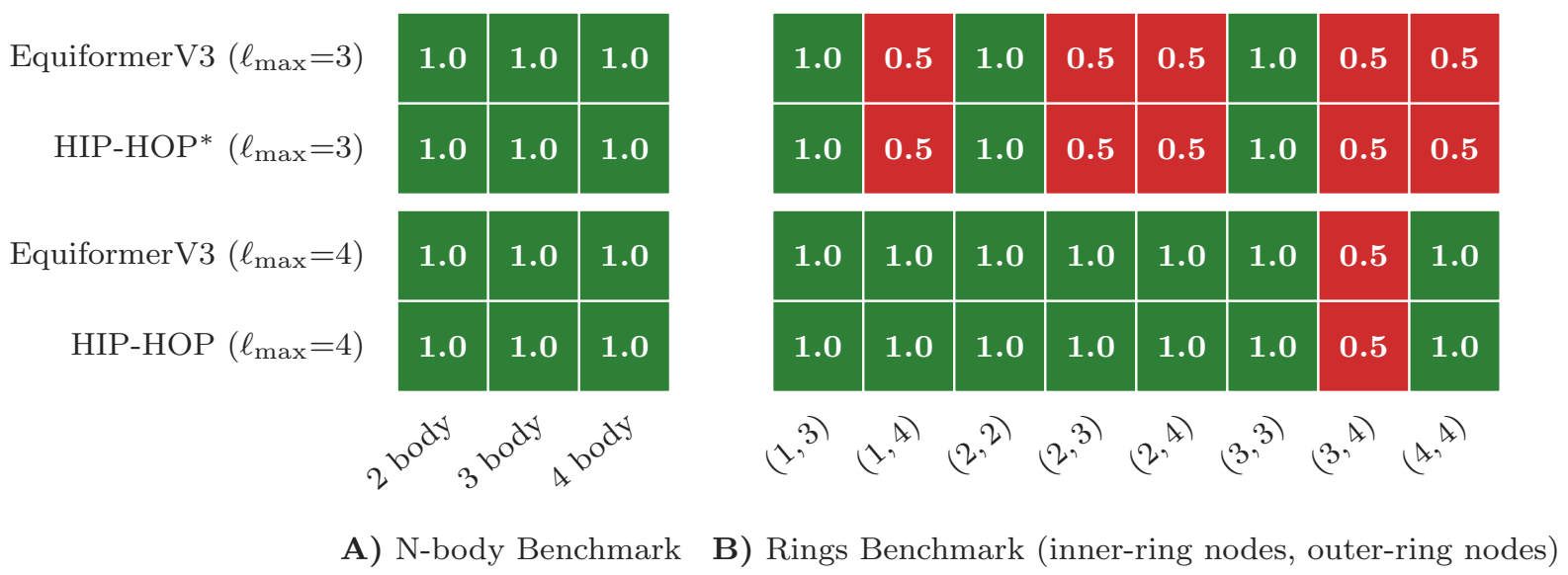
\paragraph{N-body Benchmark}

Consistent with the theoretical predictions of \cite{allen2026optimal}, HIP-HOP with basis functions up to \lthree\ successfully resolves interactions up to three-body and the expanded \lfour\ basis does not improve the expressiveness (\cref{fig:ring-nbody-results}). The static performance across all models and invariant bases motivates us to additionally evaluate with the rings benchmark to test changes in expressiveness.

\paragraph{Rings Benchmark} On the rings benchmark, the original HIP-HOP basis fails, as expected, to distinguish patterns involving four-fold symmetric configurations. Extending to \lfour\ resolves most cases; however, both HIP-HOP and EquiformerV3 fail on the (3/4) configuration (\cref{fig:ring-nbody-results}). This behavior appears consistent with the singular degeneracy recently described in \cite{domina2026}, which we discuss further in Appendix~\ref{apd:singularity}. Adding an additional contraction to the basis resolves these special configurations, though it did not provide measurable benefits in subsequent experiments (Appendix~\ref{apd:methane-details}).

\paragraph{Methane Dataset}

The new \lfour\ basis consistently achieves the lowest errors among the small HIP-HOP configurations across all four metrics (\cref{tab:methane-speed-results} and Appendix~\ref{apd:methane-details}). At roughly 36K parameters, it nearly matches EquiformerV3 in accuracy while running $3.2\times$ faster at inference. At 1.1M parameters, EquiformerV3 achieves lower error, but HIP-HOP remains $2.4\times$ faster at inference. Parallelization further improves both the one-time basis construction and inference speed, making the new basis $1.57\times$ faster (Appendix~\ref{apd:speedup-details}).
















\begin{table*}
\floatconts
{tab:methane-results}
{\caption{
The largest (\lfour, \nfour) basis has the lowest prediction error of the HIP-HOP variants while also being significantly faster than EquiformerV3.
}}
{%
\centering

\vspace{-0.75em}

\footnotesize
\setlength{\tabcolsep}{4pt}

\begin{tabular}{@{}lccccccccc@{}}
\toprule

\multicolumn{4}{c}{\bfseries Model / Basis}
& \multicolumn{4}{c}{\bfseries Prediction Performance}
& \multicolumn{2}{c}{\bfseries Inference Speed} \\

\cmidrule(lr){1-4}
\cmidrule(lr){5-8}
\cmidrule(lr){9-10}

\bfseries Model
& $\boldsymbol{\ell_{\max}}$
& $\boldsymbol{n_{\max}}$
& \bfseries \# Par.
& \bfseries E MAE
& \bfseries E RMSE
& \bfseries F MAE
& \bfseries F RMSE
& $\boldsymbol{\mu}$\bfseries s/atom
& \bfseries Speedup \\

\midrule

HIP-HOP
& 3 & 4 & 25 K
& $0.21{\scriptstyle \pm 0.02}$
& $0.50{\scriptstyle \pm 0.35}$
& $0.61{\scriptstyle \pm 0.03}$
& $2.63{\scriptstyle \pm 2.56}$
& 2.28
& 1.07$\times$
\\



HIP-HOP
& 4 & 3 & 24 K
& $0.21{\scriptstyle \pm 0.00}$
& $0.38{\scriptstyle \pm 0.02}$
& $0.62{\scriptstyle \pm 0.01}$
& $1.54{\scriptstyle \pm 0.33}$
& 3.38
& 1.05$\times$
\\

HIP-HOP
& 4 & 4 & 36 K
& $\boldsymbol{0.18{\scriptstyle \pm 0.00}}$
& $0.32{\scriptstyle \pm 0.01}$
& $0.55{\scriptstyle \pm 0.01}$
& $\boldsymbol{1.24{\scriptstyle \pm 0.03}}$
& 4.42
& 1.57$\times$
\\


EquiformerV3
& 4 & \textit{N/A} & 36 K  
& $\textbf{0.18} {\scriptstyle \pm \textit{N/A}}$
& $\textbf{0.29} {\scriptstyle \pm \textit{N/A}}$
& $\textbf{0.51} {\scriptstyle \pm \textit{N/A}}$
& $1.25 {\scriptstyle \pm \textit{N/A}}$
& 14.3
& \textit{N/A}
\\

\cmidrule(lr){1-10}

  HIP-HOP
  & 4 & 4 & 1.13 M
  & $0.04 {\scriptstyle \pm \textit{N/A}}$
  & $0.11 {\scriptstyle \pm \textit{N/A}}$
  & $0.17 {\scriptstyle \pm \textit{N/A}}$
  & $0.61 {\scriptstyle \pm \textit{N/A}}$
  & 12.7
  & 1.72$\times$ 
  \\

  EquiformerV3
  & 4 & \textit{N/A} & 1.12 M
  & $\textbf{0.03} {\scriptstyle \pm \textit{N/A}}$
  & $\textbf{0.05} {\scriptstyle \pm \textit{N/A}}$
  & $\textbf{0.10} {\scriptstyle \pm \textit{N/A}}$
  & $\textbf{0.35} {\scriptstyle \pm \textit{N/A}}$
  & 30.2
  & \textit{N/A}
  \\

\bottomrule
\end{tabular}
}
\label{tab:methane-speed-results}
\end{table*}

\vspace{-1em}
\section{Discussion and Conclusion}
\label{sec:discussion}
We introduced a benchmark for systematically identifying failures of invariant and equivariant representations in highly symmetric atomic environments. Using this benchmark, we showed that low-order representations fail to distinguish several geometrically distinct configurations, whereas increasing $\ell_{\max}$ and introducing targeted higher-order contractions can recover much of the missing geometric information.

The rings benchmark suggests that increasing angular resolution can recover geometric distinctions inaccessible to lower-order invariant bases. The failure on the (3/4)-rings configuration shows that finite invariant representations can retain isolated degeneracies even at higher $\ell_{\max}$. Interestingly, the additional contractions designed to resolve these special cases did not improve performance on the methane dataset, indicating that formal expressivity and practical predictive accuracy are not always directly aligned.

We then investigated whether the additional expressivity translates to improved molecular modeling on the methane dataset. The larger basis consistently achieved the lowest energy and force errors among the evaluated HIP-HOP configurations. 
At 36K parameters, HIP-HOP nearly matched the accuracy of EquiformerV3 while running $3.2\times$ faster at inference. At 1.1M parameters, EquiformerV3 was more accurate, though HIP-HOP remained more than twice as fast.
Together with the reduced basis-construction runtimes provided by our parallel implementation, these results demonstrate a favorable trade-off between predictive accuracy and computational efficiency.

Overall, our results provide a practical framework for diagnosing representation failures, constructing more expressive invariant bases, and evaluating the relationships among formal expressivity, predictive accuracy, and computational efficiency in machine-learned interatomic potentials.

\paragraph{Acknowledgment} We thank Kei Davis for feedback on the manuscript. This work is published under LANL LA-UR-26-27280. We gratefully acknowledge the support of the U.S. Department of Energy through the LANL Laboratory Directed Research Development Program under project number 20250145ER. 

\bibliography{bibliography}
\newpage

\appendix

\section{Methane Training Details}
\label{apd:methane-details}

We train each model using a fixed pool of $10^6$ methane configurations and evaluate performance on a separate held-out test set of 80,000 configurations.  Energies and forces are converted to kcal/mol and kcal/mol/\AA, respectively, and models are optimized using the sum of energy and force MAE and RMSE with L2 regularization. Training uses early stopping based on validation energy MAE with a patience of 300 epochs, while the batch size is increased after 150 epochs without improvement.


\begin{table}[ht]
\centering
\caption{EquiformerV3 hyperparameters used for methane training.}
\label{tab:methane-equi-hyperparameters}
\small
\begin{tabular}{lcc}
\toprule
\textbf{Hyperparameter} & \textbf{36.2 K model} & \textbf{1.12 M model} \\
\midrule
Message-passing layers & 1 & 1 \\
Hidden dimension & 16 & 32 \\
$\ell_{\max}$ & 4 & 4 \\
$m_{\max}$ & $\ell_{\max}$ & $\ell_{\max}$ \\
Attention heads & 2 & 8 \\
Attention hidden dimension & 4 & 32 \\
FFN hidden dimension & 16 & 128 \\
Edge hidden dimension & 8 & 64 \\
Radial basis functions & 20 & 20 \\
Maximum neighbors & 4 & 4 \\
Cutoff radius & 10.3 \AA & 10.3 \AA \\
\midrule
Optimizer & Adam & Adam \\
Batch size & 4096 & 256 \\
Learning rate & $2.5 \times 10^{-3}$ & $1 \times 10^{-4}$ \\
Gradient-norm clipping & 100 & 100 \\
L2 regularization & $1 \times 10^{-6}$ & $1 \times 10^{-6}$ \\
LR reduction factor & 0.5 & 0.5 \\
LR scheduler patience & 150 epochs & 150 epochs \\
Early stopping patience & 300 epochs & none \\
Max training epochs & 10,000 & 10,000 \\
\bottomrule
\end{tabular}
\end{table}

\begin{table}[ht]
\centering
\caption{HIP-HOP-NN hyperparameters used for methane training.}
\label{tab:methane-hiphop-hyperparameters}
\small
\begin{tabular}{lcc}
\toprule
\textbf{Hyperparameter} & \textbf{$\sim$25--36 K models} & \textbf{1.13 M model} \\
\midrule
Message-passing layers & 1 & 1 \\
Hidden dimension & 32 & 192 \\
Cutoff radius & 10.3 \AA & 10.3 \AA \\
\midrule
Minimum batch size & 256 & 4096 \\
Maximum batch size & 4096 & 4096 \\
Batch size doubling patience & 150 epochs & -- \\
Optimizer & Adam & Adam \\
Learning rate & $2.5 \times 10^{-3}$ & $2.5 \times 10^{-3}$ \\
LR reduction factor & -- & 0.5 \\
LR scheduler patience & -- & 150 epochs \\
L2 regularization & $1 \times 10^{-6}$ & $1 \times 10^{-6}$ \\
Early stopping patience & 300 epochs & 300 epochs \\
Max training epochs & 10,000 & 10,000 \\
\bottomrule
\end{tabular}
\end{table}


 \begin{table*}[ht]
  \floatconts
    {tab:model-performance}
    {\caption{Energy and force prediction performance for HIP-HOP-NN
  and EquiformerV3 on the methane dataset.}}
    {%
      \centering
      \resizebox{0.8\textwidth}{!}{%
        \begin{tabular}{ccccccc}
          \toprule
          \multicolumn{2}{c}{\bfseries Basis}
          &
          & \multicolumn{2}{c}{\bfseries Energy}
          & \multicolumn{2}{c}{\bfseries Forces} \\
          \cmidrule(lr){1-2}
          \cmidrule(lr){4-5}
          \cmidrule(lr){6-7}

          $\boldsymbol{\ell_{\max}}$
          & $\boldsymbol{n_{\max}}$
          & \bfseries \# Par.
          & \bfseries MAE
          & \bfseries RMSE
          & \bfseries MAE
          & \bfseries RMSE \\
          \midrule

          \multicolumn{7}{l}{\bfseries HIP-HOP-NN} \\
          \addlinespace[2pt]

          3 & 4 & 24.7 K
            & $0.213 \pm 0.016$
            & $0.500 \pm 0.347$
            & $0.607 \pm 0.032$
            & $2.631 \pm 2.557$ \\

          3 & 5 & 25.8 K
            & $0.225 \pm 0.031$
            & $0.389 \pm 0.046$
            & $0.642 \pm 0.046$
            & $3.842 \pm 4.498$ \\

          4 & 3 & 24.4 K
            & $0.213 \pm 0.004$
            & $0.378 \pm 0.024$
            & $0.623 \pm 0.013$
            & $1.537 \pm 0.333$ \\

          4 & 4 & 36.3 K
            & $0.184 \pm 0.003$
            & $0.322 \pm 0.006$
            & $0.551 \pm 0.011$
            & $1.244 \pm 0.026$ \\

          \addlinespace[2pt]
          4 & 4 & 1.13 M
            & 0.041
            & 0.107
            & 0.174
            & 0.607 \\

          \addlinespace[3pt]
          \midrule
          \addlinespace[3pt]

          \multicolumn{7}{l}{\bfseries EquiformerV3} \\
          \addlinespace[2pt]

          4 & \textit{N/A} & 36.2 K
            & 0.176
            & 0.291
            & 0.511
            & 1.248 \\

          \addlinespace[2pt]
          4 & \textit{N/A} & 1.12 M
            & 0.026
            & 0.055
            & 0.102
            & 0.351 \\

          \bottomrule
        \end{tabular}%
      }%
    }
    \label{tab:methane_results}
  \end{table*}

\newpage 
\section{N-body Details}
\label{apd:n-body-details}

Each counterexample is a pair of star-shaped local environments centered on a single node with binary labels $\{0,1\}$. We restrict message passing to only be allowed from the center node to the leaf nodes (leaf--leaf messages are blocked), making the distance cutoff irrelevant. Random accuracy is 0.5. We report performance over $2$ seeds.

  \begin{table}[ht]
  \centering
  \caption{HIP-HOP-NN hyperparameters used for the n-body benchmark.}
  \label{tab:nbody-hyperparameters}
  \small
  \begin{tabular}{lc}
  \toprule
  \textbf{Hyperparameter} & \textbf{Value} \\
  \midrule
  Message-passing layers & 1 \\
  Hidden dimension & 32 \\
  Sensitivity functions & 32 \\
  Atom layers per interaction & 2 \\
  Loss & binary cross-entropy \\
  Optimizer & Adam \\
  Learning rate & $1 \times 10^{-3}$ \\
  \bottomrule
  \end{tabular}
  \end{table}

\section{Ring Dataset Details}
\label{apd:ring-details}

\begin{figure}[ht]
    \centering

    \begin{minipage}[t]{0.49\columnwidth}
        \centering
        \includegraphics[width=\linewidth]
        {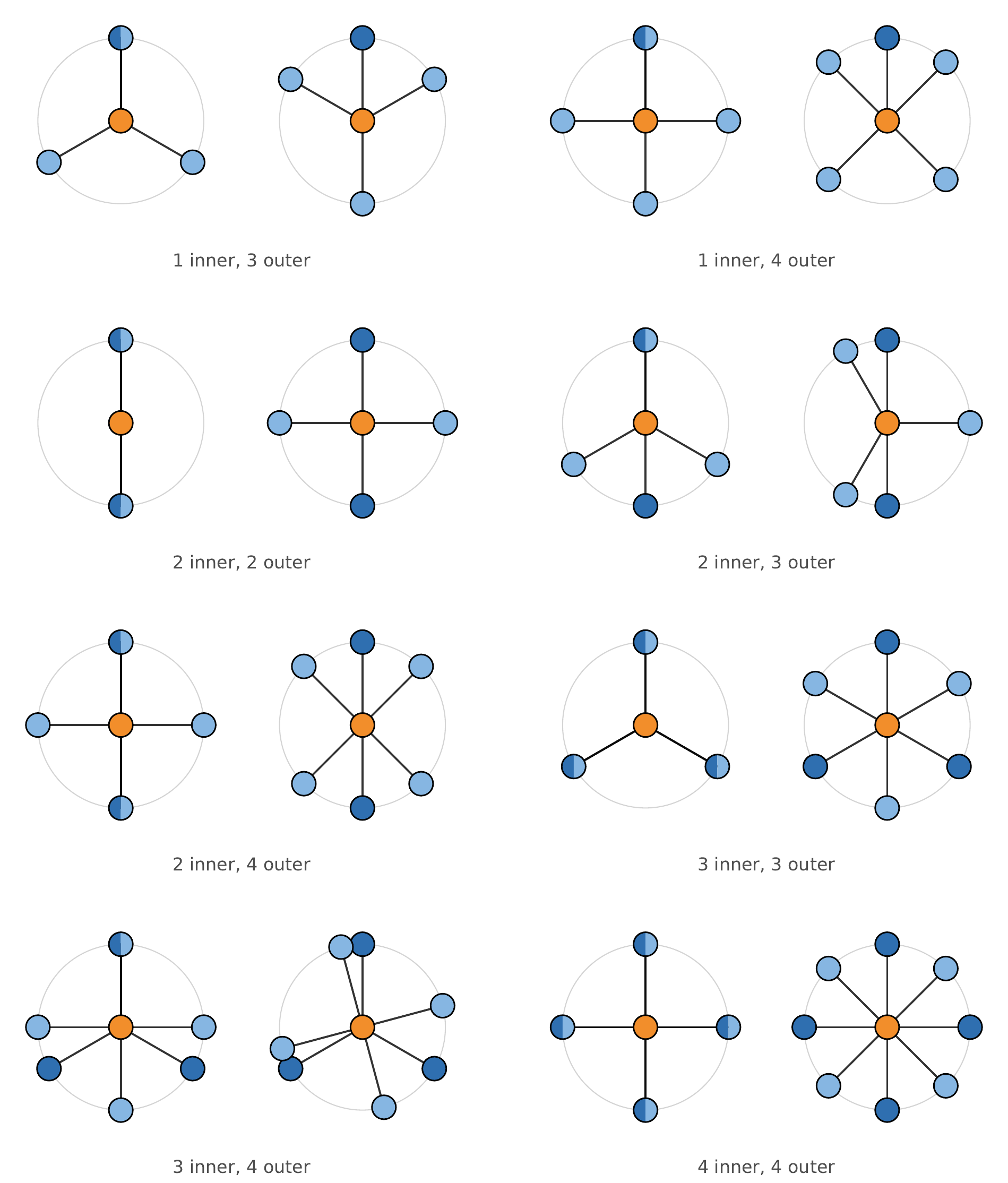}

        \small (a) Equal inner and outer radii
    \end{minipage}
    \hfill
    \begin{minipage}[t]{0.49\columnwidth}
        \centering
        \includegraphics[width=\linewidth]
        {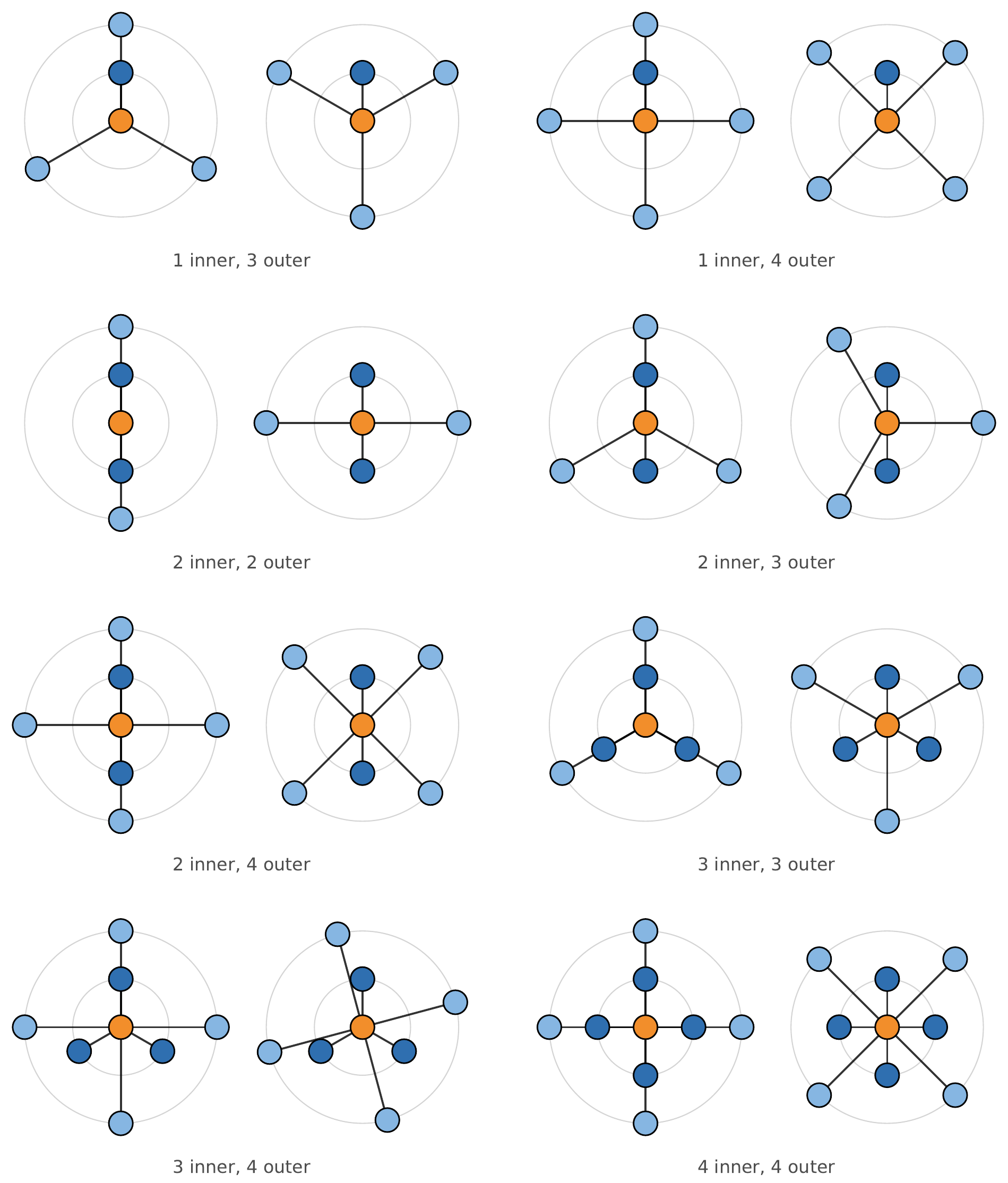}

        \small (b) Different inner and outer radii
    \end{minipage}

    \caption{
        Ring benchmark configurations with
        (a) equal radii for the inner and outer rings and
        (b) different radii for the inner and outer rings.
    }
    \label{fig:rings-samples}
\end{figure}

For each ring configuration $(n_{\mathrm{inner}},n_{\mathrm{outer}})$, we construct a local environment consisting of a central node and two rings containing $n_{\mathrm{inner}}$ and $n_{\mathrm{outer}}$ nodes. All nodes have the same species. Edges connect the central node to every ring node, with no edges between nodes on either ring. For each configuration, we generate two graphs that differ only in the distance between the rings. The graphs are assigned binary labels according to the minimum Euclidean distance between an inner-ring and outer-ring node, with the smaller-distance configuration labeled ``close'' and the larger-distance configuration labeled ``far.'' Because each task contains one graph from each class, random accuracy is 0.5. We experimented with assigning the same radius for the inner and outer rings, but found that the accuracy of the models was identical to when the radii of the rings was different (\cref{fig:rings-samples}).

\begin{table}[ht] \centering \caption{HIP-HOP-NN hyperparameters used for the ring benchmark.} \label{tab:ring-hyperparameters} \small \begin{tabular}{lc} \toprule \textbf{Hyperparameter} & \textbf{Value} \\ \midrule Graphs per configuration & 2 \\ Message-passing layers & 1 \\ Hidden dimension & 32 \\ Sensitivity functions & 32 \\ Atom layers per interaction & 2 \\ Loss & binary cross-entropy \\ Optimizer & Adam \\ Learning rate & $1 \times 10^{-3}$ \\ Max training epochs & 500 \\ \bottomrule \end{tabular} \end{table}

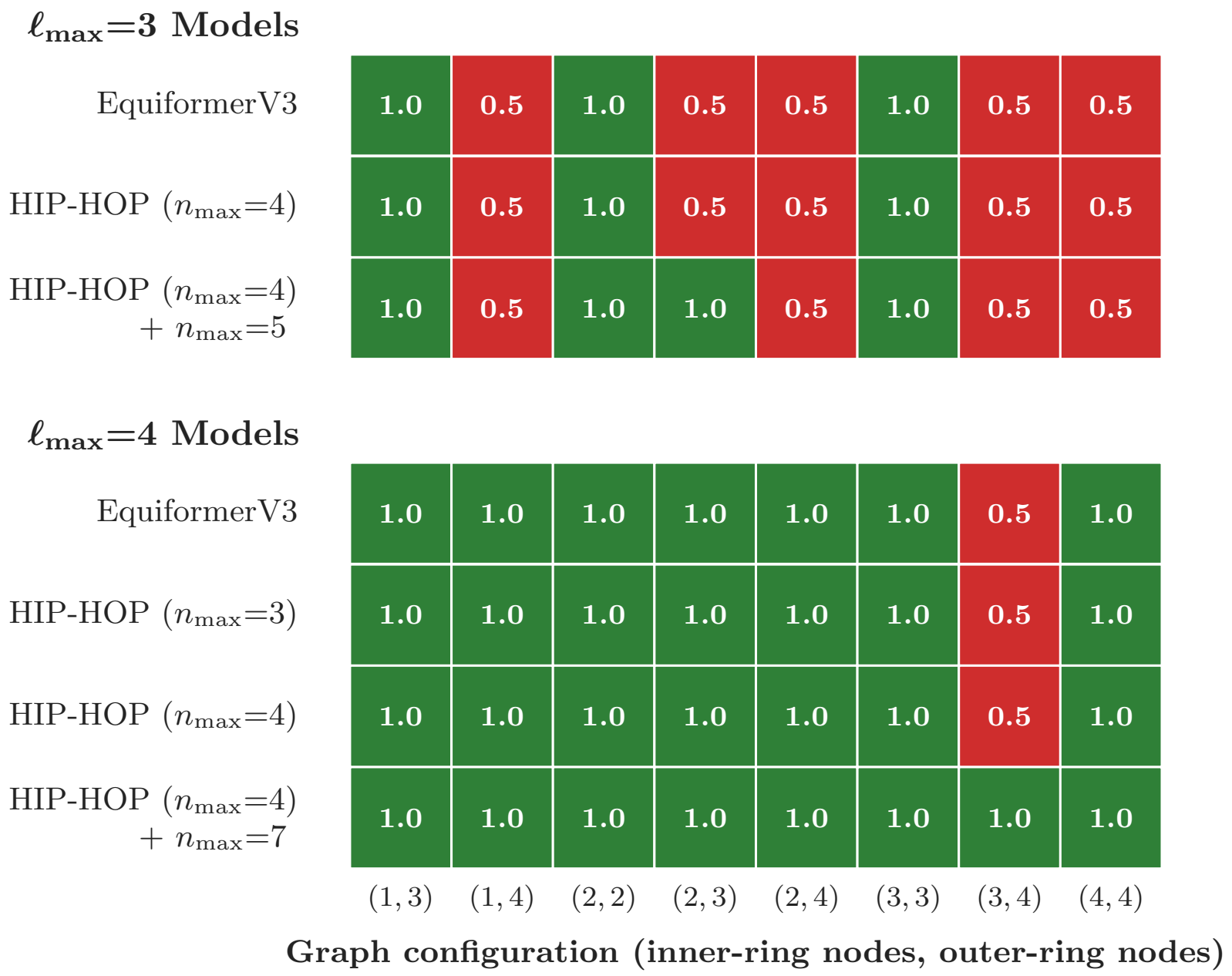
\begin{figure}[t]
\centering
\resizebox{0.7\columnwidth}{!}{%
\begin{tikzpicture}[
    x=1cm,
    y=1cm,
    font=\sffamily,
    >=Latex
]


\def\cellw{1.008}
\def\cellh{1.008}
\def\matrixw{8.064}

\def\labelx{-0.38}
\def\lmaxx{-1.85}
\def\sublineoffset{3.98em}
\def\sublinegap{-0.2ex}

\def\sectiongap{0.25}

\newcommand{\drawscorerow}[3]{%
    \foreach \score [count=\column from 0] in {#3}{%

        \ifdim\score pt>0.75pt
            \def\cellcolor{scoregood}%
        \else
            \def\cellcolor{scorelow}%
        \fi

        \fill[\cellcolor]
            ({\column*\cellw},{#1-\cellh/2})
            rectangle
            ({(\column+1)*\cellw},{#1+\cellh/2});

        \draw[
            white,
            line width=0.65pt
        ]
            ({\column*\cellw},{#1-\cellh/2})
            rectangle
            ({(\column+1)*\cellw},{#1+\cellh/2});

        \node[
            text=white,
            font=\scriptsize\bfseries
        ] at ({(\column+0.5)*\cellw},#1)
            {\score};
    }

    \node[
        anchor=east,
        text=plotink,
        font=\footnotesize
    ] at (\labelx,#1)
        {#2};
}


\node[
    anchor=center,
    text=plotink,
    font=\small\bfseries\boldmath
] at (\lmaxx,0.30)
    {\lthree\ Models};

\drawscorerow{-0.48}
    {EquiformerV3}
    {1.0,0.5,1.0,0.5,0.5,1.0,0.5,0.5}

\drawscorerow{-1.488}
    {HIP-HOP (\nfour)}
    {1.0,0.5,1.0,0.5,0.5,1.0,0.5,0.5}

\drawscorerow{-2.496}
    {\shortstack[l]{
        HIP-HOP (\nfour)\\[\sublinegap]
        \hspace*{\sublineoffset}+ \nfive
    }}
    {1.0,0.5,1.0,1.0,0.5,1.0,0.5,0.5}

\draw[
    white,
    line width=0.8pt
] (0,-0.984) -- (\matrixw,-0.984);


\node[
    anchor=center,
    text=plotink,
    font=\small\bfseries\boldmath
] at (\lmaxx,{-3.504-\sectiongap})
    {\lfour\ Models};

\drawscorerow{-4.284-\sectiongap}
    {EquiformerV3}
    {1.0,1.0,1.0,1.0,1.0,1.0,0.5,1.0}

\drawscorerow{-5.292-\sectiongap}
    {HIP-HOP (\nthree)}
    {1.0,1.0,1.0,1.0,1.0,1.0,0.5,1.0}

\drawscorerow{-6.300-\sectiongap}
    {HIP-HOP (\nfour)}
    {1.0,1.0,1.0,1.0,1.0,1.0,0.5,1.0}

\drawscorerow{-7.308-\sectiongap}
    {\shortstack[l]{
        HIP-HOP (\nfour)\\[\sublinegap]
        \hspace*{\sublineoffset}+ \nseven
    }}
    {1.0,1.0,1.0,1.0,1.0,1.0,1.0,1.0}

\draw[
    white,
    line width=0.8pt
] (0,{-4.788-\sectiongap})
  -- (\matrixw,{-4.788-\sectiongap});

\foreach \inner/\outer [count=\column from 0] in {
    1/3,
    1/4,
    2/2,
    2/3,
    2/4,
    3/3,
    3/4,
    4/4
}{%
    \node[
        text=plotink,
        font=\scriptsize
    ] at (
        {(\column+0.5)*\cellw},
        {-8.112-\sectiongap}
    )
        {$\left(\inner,\outer\right)$};
}

\node[
    text=plotink,
    font=\footnotesize\bfseries
] at (4.032,{-8.664-\sectiongap})
    {Graph configuration
    (inner-ring nodes, outer-ring nodes)};

\end{tikzpicture}%
}
\caption{
Ring benchmark performance across graph configurations for different model and invariant-basis configurations.
}
\label{fig:ring-results}
\end{figure}

\clearpage
\section{Speedups}
\label{apd:speedup-details}

Invariant-construction time was measured on a single NVIDIA Tesla V100 PCIe 32 GB GPU. Inference speed was measured with a batch size of 128 on an NVIDIA GeForce RTX 3090 GPU. For each model, the times were averaged over 4 runs after 1 warm up run. 

\begin{table}[ht]
\centering
\caption{Inference speed comparison between HIP-HOP and EquiformerV3 models at different model sizes.}
\label{tab:runtime-speedup}

\resizebox{0.75\textwidth}{!}{%
\begin{tabular}{@{}lcccccc@{}}
  \toprule

  \multicolumn{4}{c}{\bfseries Model / Basis}
  & \multicolumn{3}{c}{\bfseries Inference Speed ($\mu$s/atom)} \\

  \cmidrule(lr){1-4}
  \cmidrule(lr){5-7}

  \bfseries Model
  & $\boldsymbol{\ell_{\max}}$
  & $\boldsymbol{n_{\max}}$
  & \bfseries \# Par.
  & \bfseries Baseline
  & \bfseries Optimized
  & \bfseries Speedup \\

  \midrule

  \multirow{2}{*}{HIP-HOP}
  & \multirow{2}{*}{3}
  & \multirow{2}{*}{4}
  & 24.7 K
  & 2.45 & 2.28
  & \textcolor{speedupgreen}{\bfseries 1.07$\times$} \\

  &
  &
  & 1.02 M
  & 10.19 & 9.66
  & \textcolor{speedupgreen}{\bfseries 1.05$\times$} \\

  \cmidrule(lr){1-7}

  \multirow{2}{*}{HIP-HOP}
  & \multirow{2}{*}{3}
  & \multirow{2}{*}{5}
  & 25.8 K
  & 3.13 & 2.95
  & \textcolor{speedupgreen}{\bfseries 1.06$\times$} \\

  &
  &
  & 1.05 M
  & 16.04 & 15.76
  & \textcolor{speedupgreen}{\bfseries 1.02$\times$} \\

  \cmidrule(lr){1-7}

  \multirow{1}{*}{EquiformerV3}
  & \multirow{1}{*}{3}
  & \multirow{1}{*}{\textit{N/A}}
  & 23.1 K
  & 11.3 &  &  \\


  \midrule

  \multirow{2}{*}{HIP-HOP}
  & \multirow{2}{*}{4}
  & \multirow{2}{*}{3}
  & 24.4 K
  & 3.55 & 3.38
  & \textcolor{speedupgreen}{\bfseries 1.05$\times$} \\

  &
  &
  & 0.99 M
  & 13.90 & 13.31
  & \textcolor{speedupgreen}{\bfseries 1.04$\times$} \\

  \cmidrule(lr){1-7}

  \multirow{2}{*}{HIP-HOP}
  & \multirow{2}{*}{4}
  & \multirow{2}{*}{4}
  & 36.3 K
  & 6.93 & 4.42
  & \textcolor{speedupgreen}{\bfseries 1.57$\times$} \\

  &
  &
  & 1.13 M
  & 21.76 & 12.66
  & \textcolor{speedupgreen}{\bfseries 1.72$\times$} \\


  \cmidrule(lr){1-7}

  \multirow{2}{*}{HIP-HOP}
  & \multirow{2}{*}{4}
  & \multirow{2}{*}{7}
  & 37.4 K
  & 49.88 & 26.97
  & \textcolor{speedupgreen}{\bfseries 1.85$\times$} \\

  &
  &
  & 1.39 M
  & 318.70 & 202.54
  & \textcolor{speedupgreen}{\bfseries 1.57$\times$} \\

  \cmidrule(lr){1-7}

  \multirow{2}{*}{EquiformerV3}
  & \multirow{2}{*}{4}
  & \multirow{2}{*}{\textit{N/A}}
  & 36.2 K
  & 14.29 &  &  \\

  &
  &
  & 1.12 M
  & 30.20 &  &  \\

  \bottomrule
\end{tabular}%
}

\end{table}

\begin{table}[ht]
\floatconts
{tab:epoch-time}
{\caption{Average training time per epoch on 1M methane configurations at a matched parameter count, averaged over 5 training-only epochs on one RTX 3090. HIP-HOP trains $1.66\times$ faster than EquiformerV3.}}
{%
\centering

\vspace{-0.75em}

\footnotesize
\setlength{\tabcolsep}{4pt}

\begin{tabular}{@{}lcc@{}}
  \toprule

  \bfseries Model
  & \bfseries Params
  & \bfseries Time / Epoch (s) \\

  \midrule

  EquiformerV3 (\lfour)
  & 1.12M
  & 388.8 \\

  HIP-HOP (\lfour, $n_{\max}=4$)
  & 1.13M
  & 234.8 \\

  \bottomrule
\end{tabular}
}
\label{tab:epoch-time}
\end{table}

\begin{table}[ht]
\floatconts
{tab:invariant-construction-speedup}
{\caption{Parallelizing the construction of polynomial invariants substantially reduces construction time, especially for the larger \lfour\ bases.}}
{%
\centering

\vspace{-0.75em}

\footnotesize
\setlength{\tabcolsep}{4pt}

\begin{tabular}{@{}lccccc@{}}
  \toprule

  \multicolumn{2}{c}{\bfseries Basis}
  & \multicolumn{3}{c}{\bfseries Invariant Construction (seconds)} \\

  \cmidrule(lr){1-2}
  \cmidrule(lr){3-5}

  $\boldsymbol{\ell_{\max}}$
  & $\boldsymbol{n_{\max}}$
  & \bfseries Baseline
  & \bfseries Optimized
  & \bfseries Speedup \\

  \midrule

  3 & 4
  & 0.899 & 0.079
  & \textcolor{speedupgreen}{\bfseries 11.40$\times$} \\

  3 & 5
  & 1.791 & 0.084
  & \textcolor{speedupgreen}{\bfseries 21.35$\times$} \\

  \cmidrule(lr){1-5}

  4 & 3
  & 0.500 & 0.075
  & \textcolor{speedupgreen}{\bfseries 6.69$\times$} \\

  4 & 4
  & 4.028 & 0.138
  & \textcolor{speedupgreen}{\bfseries 29.29$\times$} \\

  4 & 7
  & 267.811 & 0.862
  & \textcolor{speedupgreen}{\bfseries 310.70$\times$} \\

  \bottomrule
\end{tabular}
}
\label{tab:invariant-construction-speedup}
\end{table}

\newpage
\section{New basis for \lfour}
\label{app:basis}

To extend HIP-HOP to higher angular resolution, we construct an invariant basis including channels up to \lfour, following the flexible moment-invariant framework of \cite{bujack2025}. 

\begin{figure}[ht]
\centering
\includegraphics[width=0.99\linewidth]{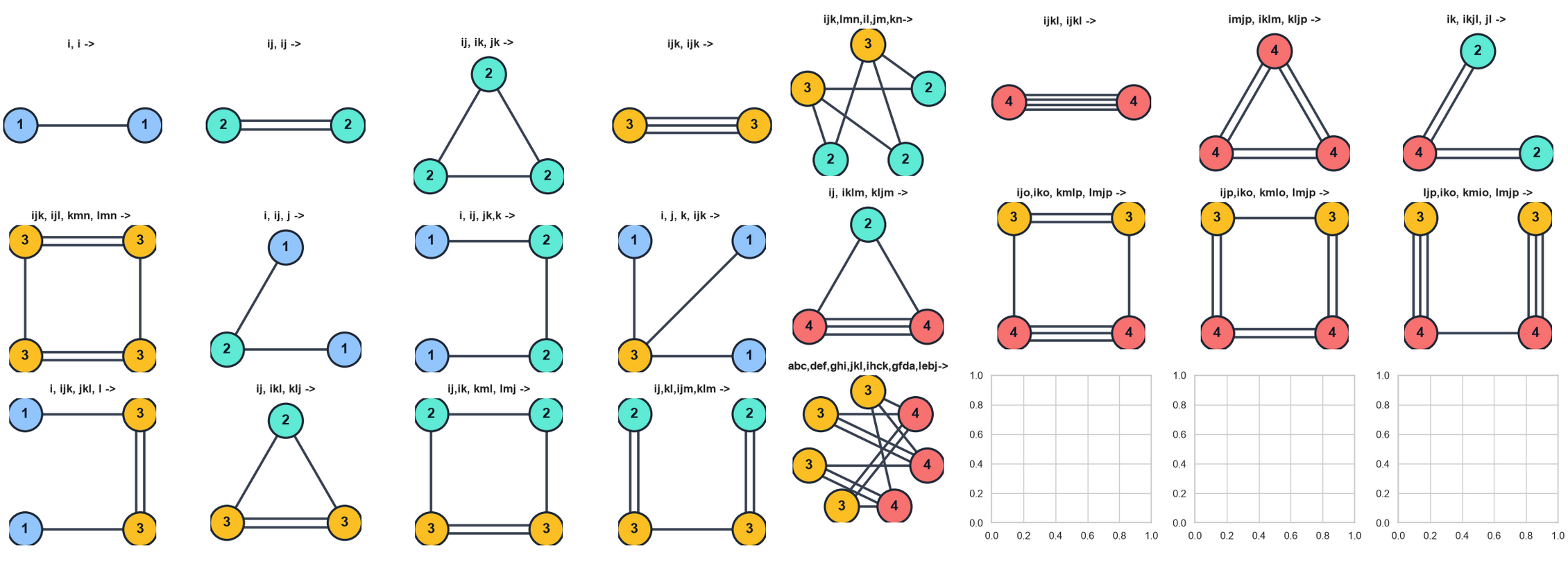}
\caption{Invariant tensor-contraction basis for HIP-HOP with angular channels up to (\lfour)}
\label{fig:l4-basis}
\end{figure}

The resulting tensor-contraction basis used in our model is summarized in Fig.~\ref{fig:l4-basis}. This figure provides a visual representation of how the different invariant scalars are calculated. Each individual graph specifies a particular tensor contraction that evaluates to a scalar. Each node labeled $\ell$ represents a rank-$\ell$ tensor $T^{(\ell)}$, while each edge represents a pair of tensor indices that are contracted with one another. Multiple edges between two nodes indicate that several indices are contracted between the same pair of tensors. The degree of each node therefore equals the rank of its corresponding tensor.

The Einstein-summation expression above each graph specifies the same contraction algebraically (\cite{weisstein_einstein_summation}). The individual comma-separated terms correspond to the nodes, while each repeated index corresponds to an edge connecting the two tensors in which that index appears. For example, the graph labeled $\texttt{i,i->}$ contains two rank-$1$ nodes joined by one edge and represents $T^{(1)}_iT^{(1)}_i$. Similarly, the graph labeled $\texttt{ij,ik,jk->}$ contains three rank-$2$ tensors. The edge corresponding to $i$ joins the first and second tensors, the edge corresponding to $j$ joins the first and third tensors, and the edge corresponding to $k$ joins the second and third tensors. 

\section{Singularity on single plane}
\label{apd:singularity}
Representing a spatial structure in a reduced or truncated basis can result in a loss of geometric information, such that distinct configurations become indistinguishable at finite $\lmax$, as recently discussed in~\cite{domina2026}. Motivated by this behavior, we arrived at a related observation independently, without the use of AI, by studying how $(n)$-fold symmetric structures behave under projection to the invariant basis. In the following, we examine this effect in the simpler setting of a single plane and characterize the problematic configurations. For simplicity, we describe the problem in the original HIP-HOP \lthree\ basis. As shown in Fig.~\ref{fig:ring-results}, the original HIP-HOP basis unexpectedly fails for the $(2/3)$-ring case (described in Fig.~\ref{fig:rings-samples}). We therefore analyze this case in more detail to understand the origin of the failure. An analogous behavior is observed for the $(3/4)$-ring configuration in the HIP-HOP $\lfour$ basis.

\begin{figure}[t]
    \centering
    \includegraphics[width=0.87\linewidth]{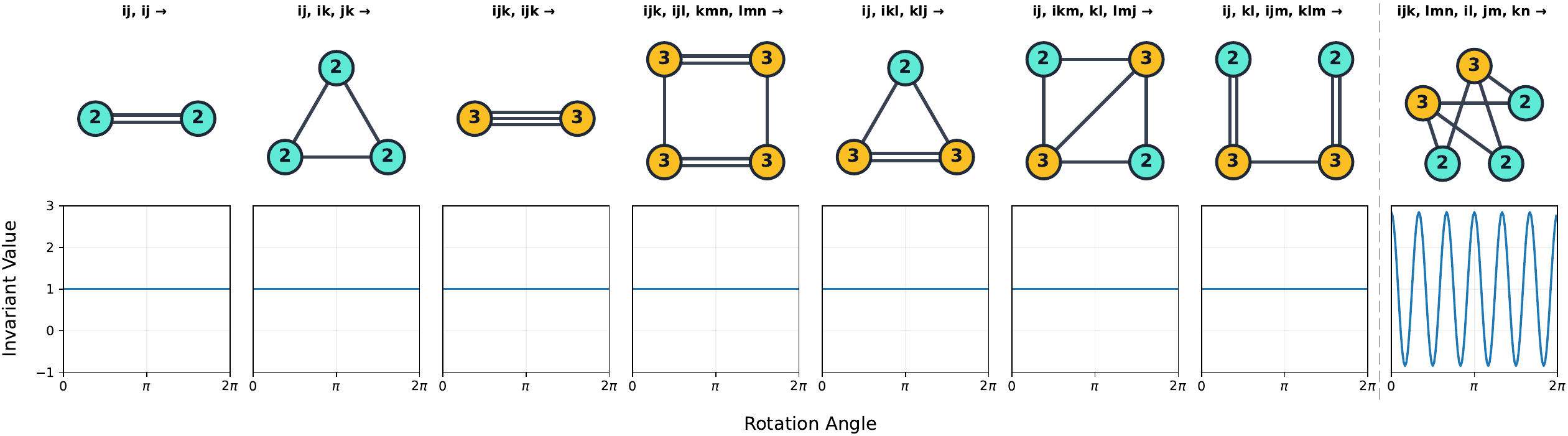}
     \caption{Nonzero contractions for the $2/3$-ring configuration as a function of the relative rotation between the two- and three-fold components. The original contractions (left) remain constant, while the additional contraction~\cite{bujack2025} (right) varies with the relative rotation.}
    \label{fig:Singularity}
\end{figure}

To analyze the $(2/3)$-ring configuration, we continuously rotate the two-fold symmetric component relative to the three-fold symmetric component and evaluate the corresponding basis functions, plotting only the nonzero responses shown in Fig.~\ref{fig:Singularity}. The original HIP-HOP basis is shown on the left, while the improved basis is shown on the right, with the two separated by a dashed line. As seen, all contractions in the original basis remain constant under the rotation, even though the configurations are distinct.  Thus, the original invariants fail to distinguish the relative orientation of the two rings. The additional contraction is functionally independent of the original contractions on this specific manifold and therefore varies with the relative rotation between the two rings. We constructed this contraction using an algorithm similar to that described in~\citep{bujack2025}.

From a tensor perspective, only the rank-2 and rank-3 tensors,
$T^{(2)}$ and $T^{(3)}$, are nonzero for this configuration.
Choosing the plane of the rings to be the $(x,y)$ plane, they can be
written as
\begin{equation}
T^{(2)}(\theta)
=
\begin{pmatrix}
a + b\cos(2\theta) & b\sin(2\theta) & 0 \\
b\sin(2\theta) & a - b\cos(2\theta) & 0 \\
0 & 0 & -2a
\end{pmatrix},
\qquad
T^{(3)}_{111} = c,
\qquad
T^{(3)}_{122}
=
T^{(3)}_{212}
=
T^{(3)}_{221}
=
-c,
\label{eq:ring_tensors}
\end{equation}
with all remaining components of $T^{(3)}$ equal to zero.
Here, $\theta$ denotes the relative rotation of the two-fold symmetric
component, while $a$, $b$, and $c$ can be arbitrary real numbers. 

Using the Einstein summation convention, the contractions are
\begin{align}
f_1
&=
T^{(2)}_{ij}T^{(2)}_{ij}
=
\operatorname{einsum}\!\left(
\texttt{ij,ij->}
\right),
\\
f_2
&=
T^{(2)}_{ij}T^{(2)}_{ik}T^{(2)}_{jk}
=
\operatorname{einsum}\!\left(
\texttt{ij,ik,jk->}
\right),
\\
f_3
&=
T^{(2)}_{ij}T^{(3)}_{ikl}T^{(3)}_{klj}
=
\operatorname{einsum}\!\left(
\texttt{ij,ikl,klj->}
\right),
\\
f_4
&=
T^{(2)}_{ij}T^{(2)}_{ik}
T^{(3)}_{kml}T^{(3)}_{lmj}
=
\operatorname{einsum}\!\left(
\texttt{ij,ik,kml,lmj->}
\right),
\\
f_5
&=
T^{(2)}_{ij}T^{(2)}_{kl}
T^{(3)}_{ijm}T^{(3)}_{klm}
=
\operatorname{einsum}\!\left(
\texttt{ij,kl,ijm,klm->}
\right),
\\
\boldsymbol{f_c}
&=
\boldsymbol{
T^{(3)}_{ijk}T^{(3)}_{lmn}
T^{(2)}_{il}T^{(2)}_{jm}T^{(2)}_{kn}
}
=
\operatorname{einsum}\!\left(
\texttt{ijk,lmn,il,jm,kn->}
\right).
\end{align}

To make the origin of the degeneracy explicit, we evaluate the basis contractions after restricting $T^{(2)}$ and $T^{(3)}$ to the forms given above. On this symmetry-constrained manifold, the tensor components depend only on $a$, $b$, $c$ and relative rotation angle $\theta$. Substituting these expressions into the contractions and simplifying them allows us to determine directly which invariants retain information about the relative orientation of the two symmetric components.

\begin{equation}
    \begin{pmatrix}
    6a^2 + 2b^2 \\
    6a(-a^2+b^2) \\
    4ac^2 \\
    4c^2(a^2+b^2) \\
    4b^2c^2 \\ 
    \boldsymbol{4 c^{2} \left(a^{3} - 3 b^{3} \sin^{2}{\left(2 \theta \right)} \cos{\left(2 \theta \right)} + b^{3} \cos^{3}{\left(2 \theta \right)}\right)}
    \end{pmatrix}
    \label{eq}
\end{equation}

None of the contractions in the original basis contains any dependence on the relative rotation angle $\theta$. Consequently, these contractions are insensitive to changes in the relative orientation of the two- and three-fold symmetric components on this manifold. Unlike the contractions in the original basis, $f_c$ retains explicit dependence on $\theta$ and therefore contains information about the relative orientation of the two symmetric components.

An analogous situation occurs for the $3/4$-ring configuration in the $\lfour$ basis, where the missing angular dependence can be recovered by the additional contraction
\begin{equation}
\operatorname{einsum}\!\left(
\texttt{abc,def,ghi,jkl,ihck,gfda,lebj->}
\right).
\end{equation}
Although these additional contractions are important from a theoretical perspective, they are computationally expensive and provide only limited practical improvement in the experiments, as shown in Table~\ref{tab:model-performance}. We interpret this as being consistent with the expectation that exactly planar atomic configurations associated with these singularities are very rare in realistic datasets.
\end{document}